\documentclass[conference,a4paper]{IEEEtran}
\IEEEoverridecommandlockouts
\usepackage{cite}
\usepackage{amsmath,amssymb,amsfonts}
\usepackage{algorithmic}
\usepackage{graphicx}
\usepackage{textcomp}
\def\BibTeX{{\rm B\kern-.05em{\sc i\kern-.025em b}\kern-.08em
    T\kern-.1667em\lower.7ex\hbox{E}\kern-.125emX}}

\usepackage{hyperref}

\begin{document}

\begin{titlepage}
	\begin{center}
		
		\Huge
		\textbf{Kalman Filter based localization in hybrid BLE-UWB positioning system}
		
		\vspace{0.5cm}
		\LARGE
		Accepted version
		
		\vspace{1.5cm}
		
		\text{Marcin Kolakowski}
		
		\vspace{.5cm}
		\Large
		Institute of Radioelectronics and Multimedia Technology
				
			 Warsaw University of Technology
			 
			  Warsaw, Poland,
			 
			 contact: marcin.kolakowski@pw.edu.pl

		\vspace{2cm}

	\end{center}

\Large
\noindent
\textbf{Originally presented at:}

\noindent
2017 IEEE International Conference on RFID Technology \& Application (RFID-TA), Warsaw, Poland, 2017

\vspace{.5cm}
\noindent
\textbf{Please cite this manuscript as:}

\noindent
M. Kolakowski, "Kalman filter based localization in hybrid BLE-UWB positioning system," 2017 IEEE International Conference on RFID Technology \& Application (RFID-TA), Warsaw, Poland, 2017, pp. 290-293, doi: 10.1109/RFID-TA.2017.8098889.

\vspace{.5cm}
\noindent
\textbf{Full version available at:}

\noindent
\url{https://doi.org/10.1109/RFID-TA.2017.8098889}

\vspace{.5cm}
\noindent
\textbf{Additional information:}

\noindent
Additional info

		\vfill

\large
\noindent
			© 2017 IEEE. Personal use of this material is permitted. Permission from IEEE must be obtained for all other uses, in any current or future media, including reprinting/republishing this material for advertising or promotional purposes, creating new collective works, for resale or redistribution to servers or lists, or reuse of any copyrighted component of this work in other works.
\end{titlepage}

\title{Kalman Filter based localization in hybrid BLE-UWB positioning system \\
}

\author{\IEEEauthorblockN{Marcin Kolakowski}
\IEEEauthorblockA{\textit{Institute of Radioelectronics and Multimedia Technology} \\
\textit{Warsaw University of Technology}\\
Warsaw, Poland \\
m.kolakowski@ire.pw.edu.pl}
}

\maketitle

\begin{abstract}
In this paper a concept of  hybrid Bluetooth Low Energy (BLE) – Ultra-wideband (UWB) positioning system is presented. The system is intended to be energy efficient. Low energy BLE unit is used as a primary source of measurement data and for most of the time localization is calculated based on received signal strength (RSS). UWB technology is used less often. Time difference of arrival (TDOA) values measured with UWB radios are periodically used to improve RSS based localization. The paper contains a description of proposed hybrid positioning algorithm. Results of simulations and experiments confirming algorithm's efficiency are also included.
\end{abstract}


\section{Introduction}
In recent years the demand for indoor location based services (LBS) has been rising.  Currently a lot of different applications can be found. They range from professional solutions in factories to localize machines and personnel to more casual ones like location based museum guides or positioning systems intended for home use. Most of these services demand accurate location information, which is not easy to obtain in indoor environments. Therefore a lot of different techniques has been considered to supply the user with effective and reliable systems. One of them is hybrid localization.

In hybrid positioning, object localization is computed based on different type of measurement data obtained from one or few separate systems. One of the most popular hybrid systems combine inertial measurement units (IMU) with conventional radio e.g. ultra-wideband (UWB) localization systems. In these solutions data from IMU are usually used to supply the user with localization in places, where accurate positioning based solely on radio signals is hard or impossible \cite{Zwirello2013}.

Another approach to hybrid localization is the use of algorithms combining measurement results of different signal parameters for example time difference of arrival (TDOA) with received signal strength (RSS). In \cite{Lategahn2013} a hybrid Extended Kalman Filter based algorithm for UWB positioning system is proposed. In that implementation RSS results are used to identify non-line of sight (NLOS) conditions and correct the localization results obtained using TDOA. Similar solution is presented in \cite{Zhu2011}. Such approach has been also investigated for other signals e.g. WCDMA \cite{Ghannouchi2012} or Wi-Fi \cite{Kumarasiri2016}.

Solutions, in which combined data are obtained using different radio technologies are less popular. In \cite{Khan2013} an EKF based algorithm combining ranging results obtained with both UWB and ZigBee modules is presented. Particle filter presented in \cite{Peltola2016} combines foot mounted inertial, Bluetooth low energy (BLE) and UWB technologies with map matching into an indoor navigation system. Both referenced papers proved that combining accurate UWB based distance measurements with narrowband RSS results allows to improve system accuracy. In systems presented in those papers, UWB based results are supplied to the algorithm at the same rate as results obtained with BLE or ZigBee. In this paper a slightly different approach is proposed.

In the paper a concept of energy efficient BLE-UWB hybrid positioning system is presented. In the system localization is calculated with Extended Kalman Filter based algorithm, which combines RSS obtained with BLE technology with TDOAs measured using UWB radio units. Due to its lower energy consumption BLE unit is used more often than its UWB counterpart. Therefore for most of the time localization is calculated based on solely RSS results. TDOA results are obtained less often and are used to periodically improve RSS based localization. Such approach allows to maintain both high location update rate, high accuracy and is energy efficient, which might be crucial in many mobile applications.

The proposed system concept is described in section II. Extended Kalman Filter based hybrid algorithm is presented in section III. Sections III and IV include the results of conducted simulations and experiments. Section V concludes the paper.

\section{System architecture}
\begin{figure}[!t]
\centerline{\includegraphics[width=3.5in]{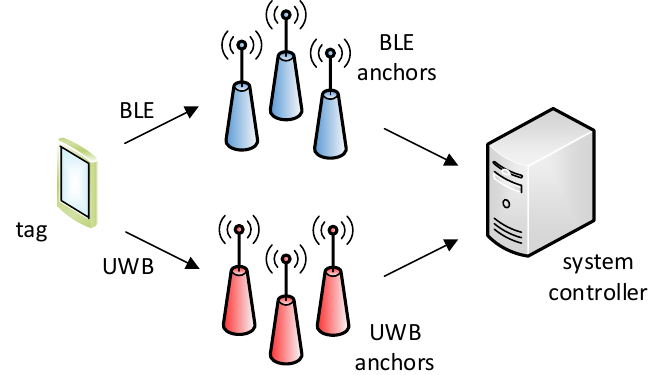}}
\caption{Hybrid localization system architecture}
\label{fig:arch}
\end{figure}

Architecture of the proposed hybrid localization system is presented in Fig.\ref{fig:arch}. The system consists of a localized device (tag), system infrastructure and a system controller.

Tag localization is calculated based on RSS and TDOA measurements conducted by the system infrastructure.  The system infrastructure comprises two types of anchor nodes: BLE anchors and UWB anchors. In the given example, they are separate devices but integrated solutions can be used as well.

The tag is a low-energy device equipped with BLE and UWB radio modules and additional motion sensors (e.g. accelerometer). It periodically sends BLE and UWB packets. Bluetooth packets are broadcasted in the advertising channel up to 10 times per second. They are used to conduct power measurements and act as data carriers for the results obtained with tag motion sensors. Ultra-wideband packets, used for time measurements are sent less often than Bluetooth packets (up to 2 times per second). Such approach allows to lower tags power consumption. Reception times of those packets are registered by UWB anchors and are used to calculate TDOA values.

Measured UWB reception times, BLE signal power levels  and sensors measurement results are sent to the system controller. There measurement data are processed, TDOA values are derived and current tag position is calculated.

\section{Localization algorithm}
The proposed localization algorithm is based on Extended Kalman Filter (EKF) \cite{grewalKalman} and fuses RSS measurement results with measured TDOA values. In the algorithm, localized tag is treated as a dynamic system, which state at the given moment $k$ is described by a state vector $x_k$  containing tag location coordinates $x, y$ and its velocity components $v_x, v_y$.
\begin{equation}
x_k = \begin{bmatrix}
x&y&v_x&v_y\\
\end{bmatrix}\label{eq:state_vec}
\end{equation}

A single EKF iteration consists of two phases: time-update and measurement update. In time-update phase current value of state vector is predicted based on the value obtained in the previous iteration and equations of motion. Time-update phase is described with the following equations:
\begin{equation}
\hat{x}_{k(-)}=F\hat{x}_{k-1(+)}\label{eq:TU}
\end{equation}
\begin{equation}
P_{k(-)}=FP_{k-1(+)}F^T+Q_k\label{eq:CTU}
\end{equation}
where $\hat{x}_{k(-)}$  is predicted state vector value, $\hat{x}_{k-1(+)}$  is state vector value obtained in the previous EKF iteration, $P_{k(-)}$ and  $P_{k-1(+)}$  are state covariance matrices of these values, $F$ is state transition matrix  containing motion equations and $Q_k$ is a process noise covariance matrix. In this implementation discrete white noise acceleration (DWNA) kinematic model was chosen. According to this model the localized object, between the analyzed moments, moves with constant speed and object acceleration is treated as process noise. Covariance  matrix $Q$ for this model is described in \cite{shalomEst}.

The predicted state vector value is updated with measurement results in measurement-update phase. In the proposed algorithm RSS and TDOA results are used. Measurement update phase is described with the set of following equations.
\begin{equation}
K_k = P_{k(-)}H_k^T\left(H_kP_{k(-)}H_k^T+R_k\right)^{-1}\label{eq:gain}
\end{equation}
\begin{equation}
\hat{x}_{k(+)}=\hat{x}_{k(-)}+K_k\left(z_k-h_k(\hat{x}_{k(-)})\right)\label{eq:MU}
\end{equation}
\begin{equation}
P_{k(+)}=\left(I-K_kH_k^T\right)P_{k(-)}\label{eq:CMU}
\end{equation}
\begin{equation}
z_k = \begin{bmatrix}
RSS_1&\cdots&RSS_n&T_1&\cdots&T_m
\end{bmatrix}\label{eq:z}
\end{equation}
\begin{equation}
\begin{split}
h_k(x_k) &=
[\begin{matrix} RSS_1(x_k)&\cdots&RSS_n(x_k) \end{matrix} \\
&\qquad\qquad\qquad\begin{matrix} T_1(x_k)&\cdots&T_m(x_k) \end{matrix}]
\end{split}
\label{eq:hz}
\end{equation}
where $z_k$ is measurement vector containing RSS ($RSS_n$) and TDOA ($T_m$)  measurement results, $h_k(x_k)$  is measurement function used to calculate measurement values which would be obtained for the predicted tag localization, $H_k$  is a linearization of that function, $K_k$  is Kalman gain and $R_k$  is measurement covariance matrix.

The length of measurement vector $z_k$  and the form of measurement function $h_k$  are not constant, because they depend on number of results, which were provided by system infrastructure. Measurement rate of RSS in the proposed system is higher than measurement rate of TDOA. Therefore, most of the time $z_k$  contains  only power measurements. To calculate predicted RSS values log-distance path loss model is used. According to this model received signal power measured by the anchor can be expressed as
\begin{equation}
RSS_n(x_k) = RSS_0 -10\gamma\log_{10}  \frac{d(x_k)}{d_0}\label{eq:rss}
\end{equation}
where $RSS_n$  is signal power received by the anchor $n$, $d$ is the distance between the anchor and predicted tag localization, $RSS_0$  is received power at the reference distance from the tag $d_0$  and $\gamma$  is path-loss exponent.

The presented algorithm allows to localize tag in two dimensions, but it can be easily extended to handle three dimensional problems.

\section{Simulations}
The proposed localization system was simulated in Matlab environment. The main goal of the simulation was to verify the system concept and algorithm effectiveness. The tests were conducted in a simulated system consisting of four BLE anchors and four UWB anchors placed on a walls of a large room. During the tests localization of a moving person was analyzed.

Measurements conducted by the system infrastructure were simulated in the following way. Received signal strength of Bluetooth signal was calculated using log-distance path loss model (\ref{eq:rss}). The reference distance $d_0$ was set to 1 m. Power received by an anchor at the reference distance equaled  40~dBm. The path-loss exponent of 1.9 was assumed, which is a value typical for indoor LOS conditions \cite{Janssen1992}. Shadowing effects which may occur in the indoor environments were taken into account by adding random zero-mean Gaussian variable with standard deviation equal 3 dB. In the simulation it was assumed that RSS measurement rate equals 10 Hz.

Time difference of arrival measurement results were derived by calculating the propagation time between the moving tag and UWB anchors and subtracting them from one another. The uncertainty of time measurements conducted by the anchors was simulated by adding to propagation times a zero mean Gaussian variable with standard deviation of 0.2ns. During the simulation stage, the effect of TDOA measurements rate on algorithm accuracy was examined. Time Difference of Arrival results were generated with different rates ranging from 1/4 to 10 Hz.

In the simulations, the localized person was moving along a trajectory consisting of straight lines. It was assumed that the person was moving with a speed of 1.4 m/s.

\begin{figure}[!t]
\centerline{\includegraphics[width=3.5 in]{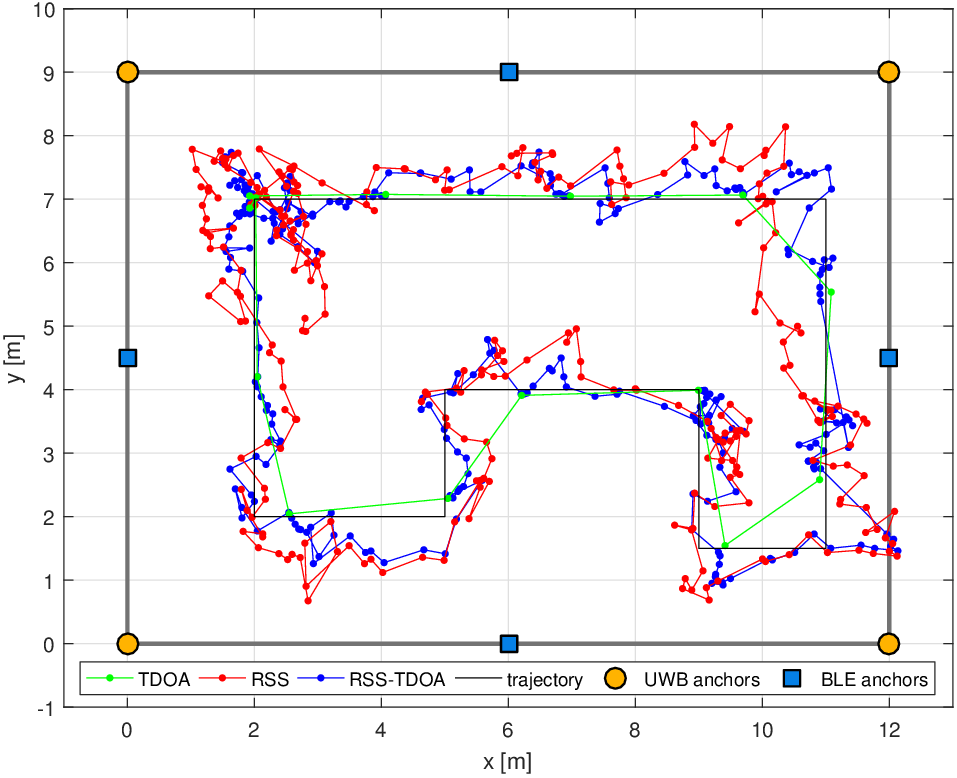}}
\caption{Simulation results. RSS rate = 10 Hz, TDOA rate = 0.5 Hz}
\label{fig:sim_res}
\end{figure}

\begin{figure}[!t]
\centerline{\includegraphics[width=3.5 in]{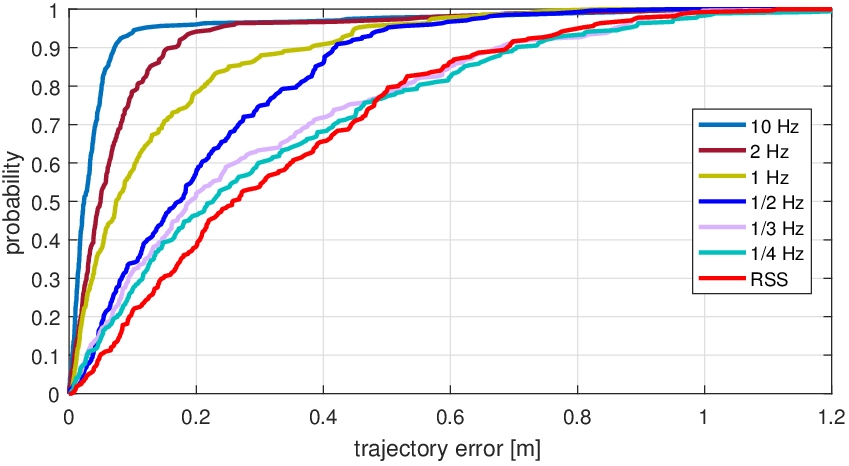}}
\caption{Empirical CDF of trajectory error for different TDOA rates}
\label{fig:sim_CDF}
\end{figure}

Tag localization was calculated using three different algorithms: EKF using only TDOA results, EKF using only RSS results and the algorithm proposed in the paper, which utilized both RSS and TDOA values. Exemplary localization results for TDOA update rate equal to 0.5 Hz are presented in Fig.\ref{fig:sim_res}. Empirical cumulative distributive function curves (CDF) of trajectory error defined as the distance between localized points and the real trajectory are presented in Fig.\ref{fig:sim_CDF}.

 Algorithm using TDOA measurement results is accurate but due to its relatively low update rate, it does not allow to precisely recreate persons movement trajectory. In case of EKF using solely RSS measurements, the obtained results are less accurate but thanks to their higher update rate it is possible to retrieve more trajectory details. 

The use of both RSS and TDOA measurement results allows to reproduce the trajectory more accurately. The additional periodic use of more accurate TDOA results  allows to reduce negative effects caused by shadowing effects. In case of using RSS results more than half of the results are located further than 26 cm from the trajectory. In case of the novel algorithm (TDOA rate 0.5 Hz) median error is lower and equals 18 cm.

The accuracy of presented algorithm depends on how often TDOA results are used. System accuracy rises with TDOA update rate. In case of highest update rate 50\% results are closer than 5 cm to the trajectory. Lowering measurement rate degrades the results. For update rates of 1/4 Hz or lower the algorithm does not significantly improve trajectory reproduction.

\section{Experiments}
The presented system concept was tested experimentally. The experiments were conducted in a fully furnished apartment.

Time difference of arrival measurements were made using DW1000 \cite{DW1000} based UWB localization system. The system infrastructure comprised six anchors which were placed in different rooms. The maximum TDOA measurement rate possible in that system was 0.16 s.

Bluetooth power levels were measured using a set of two Texas Instruments BLE evaluation boards. Multi-standard CC2650 LaunchPad \cite{CC2650} was used as a transmitter, which was set to beacon mode and transmitted 10 BLE packets per second. The transmission took place in one of the advertising channels. RSS measurements were made with CC2540 USB Evaluation Kit \cite{CC2540}. It allows to measure RSS with a resolution of 1 dB. The reported power measured at 1 m reference distance equaled -38 dBm. Path-loss exponent value of 3.3 was assumed. This value is a typical value for NLOS channels at 2.4 GHz frequency \cite{Janssen1992}.

\begin{figure}[!t]
\centerline{\includegraphics[width=3.5 in]{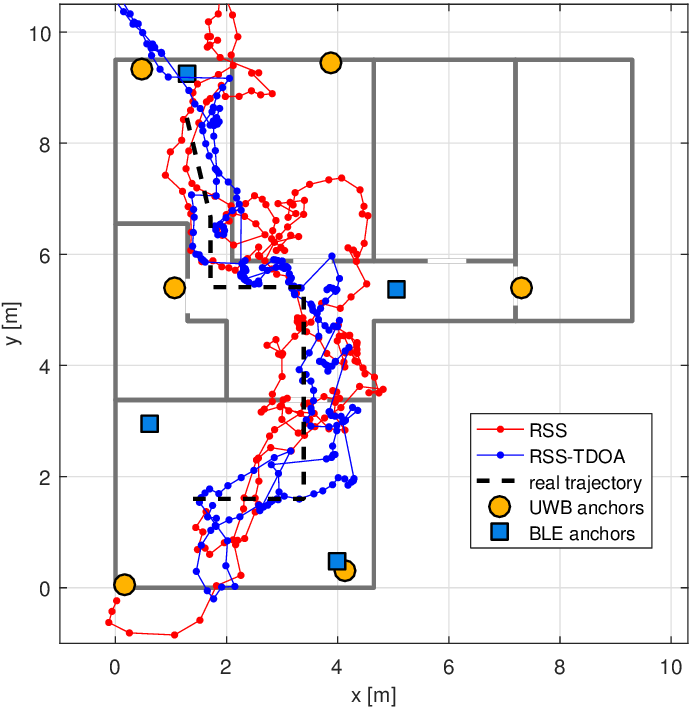}}
\caption{Localization results obtained for TDOA update rate = 0.5 Hz}
\label{fig:exp_res}
\end{figure}

\begin{figure}[!t]
\centerline{\includegraphics[width=3.5 in]{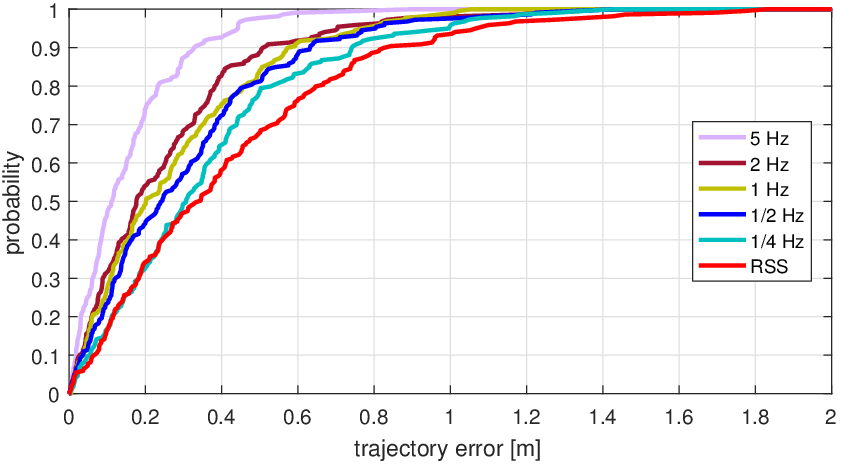}}
\caption{Empirical CDF of distance from trajectory for different TDOA measurement rates}
\label{fig:exp_CDF}
\end{figure}

During the experimental phase, a person moving along predefined trajectory was localized. Persons localization was calculated using two algorithms: EKF using solely RSS results and the algorithm proposed in the paper. Exemplary localization results for TDOA update rate equal 0.5 Hz are presented in Fig.\ref{fig:exp_res}. The empirical CDF curves of trajectory errors for different TDOA update rates are presented in Fig.\ref{fig:exp_CDF}.

Localization results calculated based on RSS results only are not very accurate. About 50\% percent of the obtained results are further than 35 cm from the  real trajectory. Using both TDOA and RSS results allows to recreate movement trajectory with higher accuracy. Just like in the simulations the level of accuracy improvement depends on TDOA measurement rate. In case when TDOA is measured with 5 Hz rate median of trajectory error has the value around 11 cm. For the TDOA update rate equal 0.5 Hz (Fig.\ref{fig:exp_CDF}), median value is around 24 cm, which is 11 cm lower than for RSS only based localization.

\section{Conclusions}
In the paper a concept of an energy efficient,  hybrid BLE UWB localization system is presented. 

The proposed positioning algorithm, based on Extended Kalman Filter, fuses RSS of Bluetooth Low Energy signals  with TDOAs obtained for UWB signals. Both signals are periodically sent by the tag, but UWB signal rate is much lower. 

 The simulations and experiments were carried out to verify system concept and positioning algorithm efficiency. Obtained test results confirm that combining both RSS and TDOA measurement results leads to positioning accuracy improvement. The quality of movement trajectory reproduction rises with TDOA measurement rate. 
 
The energy consumption of current UWB transceivers is much higher than BLE modules. The proposed solution by limiting UWB transmitter activity lowers energy requirements, but still preserves positioning accuracy. In a real system implementation TDOA measurement rate can be adapted to the tag mobility. Fast movements detected with accelerometer can trigger more frequent TDOA measurements.

\bibliography{IEEEabrv,rfidta2017_bibl}

\end{document}